\documentstyle[12pt]{article}
\begin{document}
\begin{center}
\Large\bf
Fusion Cycles in Stars and Stellar Neutrinos\\[2.6cm]
\large
G. Wolschin
\footnote{Email:  wolschin@uni-hd.de
\hspace{5cm}http://wolschin.uni-hd.de}\\[.8cm]
\normalsize\sc\rm
Universit\"at Heidelberg, D-69120 Heidelberg, Germany\\[2.0cm]
\end{center}
\bf
Abstract\\[1cm]
\rm
Starting from the early works by Weizs\"acker and Bethe about fusion cycles and
energy conversion in stars, a brief survey of thermonuclear processes in
stars leading to contemporary research problems in this field is given.
Special emphasis is put on the physics of stellar and, in particular, solar
neutrinos which is at the frontline of current
investigations.
\newpage
\section{Introduction}
In the 1920s Eddington formulated the hypothesis that fusion
reactions between light elements are the energy source of
the stars - a proposition that may be considered as the
birth of the field of nuclear astrophysics [1]. It was
accompanied by his pioneering work on stellar structure
and radiative transfer, the relation between stellar mass
and luminosity, and many other astrophysical topics.
Atkinson and Houtermans [3] showed in more detail in 1929
- after Gamow [2] had proposed the tunnel effect - that
thermonuclear reactions can indeed provide the energy source
of the stars: they calculated the probability for a nuclear
reaction in a gas with a Maxwellian velocity distribution.
In particular, they considered the penetration probability of protons
through the Coulomb barrier
into light nuclei at stellar temperatures of $4\cdot10^{7}K$.
>From the high penetration probabilities for the lightest
elements they concluded that the build-up of alpha-particles
by sequential fusion of protons
could provide the energy source of stars.
An improved formula was provided by
Gamow and Teller [4].

Hence, hydrogen and helium (which were later - in the 1950s - identified
as the main remnants of the big bang) form the basis for the synthesis
of heavier elements in stars - but details of the delicate chain reactions
that mediate these processes remained unknown
until 1938. This is in spite of the fact that
rather precise models of the late stages of stellar
evolution existed or were soon developed. At that time, white dwarfs
were generally considered to be the endpoints of stellar evolution,
although Zwicky and Baade had speculated in 1934 that a neutron star could
be the outcome of a supernova. In 1939 Oppenheimer and Volkoff presented
the first models of neutron stars as final stages of stellar evolution.
Together with the so-called "frozen" or "collapsed" stars - which were
re-named by Wheeler in 1967 as "black holes" - Chandrasekhar included these
results for sufficiently massive progenitors in his book about stellar
structure [5].
Eddington strongly rejected the proposal, but
it proved to be true when the first rotating neutron star
(pulsar) was detected in 1967 by Bell and Hewish.

Probably the most important breakthrough regarding the recognition of
fusion cycles occured in 1937/8 when Weizs\"acker [6,7] and Bethe [9]
found the CNO-cycle - which was later named after their
discoverers  Bethe-Weizs\"acker-cycle (figure 3) - in completely independent
works, and Bethe and Critchfield [8] first outlined the
proton-proton chain (figure 2). After a brief review of thermonuclear
reactions in section 2, these nucleosynthesis mechanisms are reconsidered
in section 3.

After World War II, stellar nucleosynthesis was studied further
by Fermi, Teller, Gamow, Peierls and others, but it turned out
to be difficult to understand the formation of elements heavier
than lithium-7 because there are no stable nuclei with mass
numbers 5 or 8. In 1946 Hoyle interpreted the iron-56 peak
in the relative abundances of heavier elements vs. mass (figure 1) as
being due to an equilibrium process inside stars at a
temperature of $3\cdot10^{9}K$. Later Salpeter showed
that three helium nuclei could form carbon-12 in
stars, but the process appeared to be extremely
unlikely. To produce the observed abundances, Hoyle
predicted an energy level at about 7 MeV excitation
energy in carbon-12, which was indeed discovered
experimentally, generating considerable excitement and
progress in the world of astrophysics. Burbidge, Burbidge,
Fowler and Hoyle then systematically worked out
the nuclear reactions inside stars that are the basis of
the observed abundances and summarized the field in 1957 [10].

The role of stellar  neutrinos was considered by Bethe
in [9]. Neutrinos had already been
postulated by Pauli in 1930 to interpret the continuous
beta-decay spectra, but could not be confirmed experimentally
until 1952 by Cowan and Reines. Bethe argued in 1938
that fast neutrinos emitted from beta-decay
of lithium-4 (which would result from proton capture by
helium-3) above an energy threshold of 1.9 MeV might produce
neutrons in the outer layers of a star.
However, this required the assumption of long-lived lithium-4, which turned
out to be wrong.
Bethe
did not pursue stellar neutrinos further in his early works, and he or
Weizs\"acker also did not explicitely consider
the role of neutrinos in the initial p-p reaction, or in the CNO-cycle at
that time.

In a normal star, electron neutrinos that are generated in the central region
usually leave the star without interactions that modify their energy.
Hence, the neutrino energy is treated separately from the thermonuclear
energy released by reactions, which undergoes a diffusive transport
through the stellar material that is governed by the temperature
gradient in the star. Stellar neutrinos are generated not only in nuclear
burnings
and electron capture, but also by purely leptonic processes such as
pair annihilation or Bremsstrahlung.

Neutrinos from nuclear processes in the interior of the sun should
produce a flux of $10^{11}$ neutrinos per $cm^{2}second$ on the earth.
In 1967 Davis et al. - following suggestions by Pontecorvo, and by Bahcall
and Davis
to use neutrinos ".. to see into the interior of a star and thus verify
directly the hypothesis
of nuclear-energy generation in stars" -
indeed succeeded to  measure solar neutrinos with a detector based on
390000 liters
of tetrachloroethylene. When electron-neutrinos travelling from the sun hit
the chlorine-37 nuclei, they occasionally produced argon-37 nuclei, which
were extracted
and counted by their radioactive decay. First results were published in
1968 [11].
However, these were neutrinos with higher energies produced in a side
branch of the proton-proton chain. More than 90 per cent of the neutrinos
are generated
in the initial p-p reaction, and these were observed first by the Gallex
collaboration in 1992
[12] using  gallium-71 nuclei as target in a radiochemical detector.
Together with the corresponding results of the Sage collaboration, this
confirmed experimentally the early suggestions by Weizs\"acker and Bethe
that p-p fusion is the source of solar energy.

The measurements [11,12] showed that less than 50 per cent
of the solar neutrinos that are expected to arrive on earth
are actually detected. The subsequent controversy whether this is due
to deficiencies in the solar models, or caused by flavor oscillations
was resolved at the beginning of the 21st century by combined
efforts of the SuperKamiokande [14] and SNO [15]-collaborations
in favor of the particle-physics explanation: Neutrinos have a small, but finite
mass, and hence, they can oscillate and therefore escape detection,
causing the "solar neutrino deficit" - with the size of the discrepancy
depending on energy. The identification of oscillating
solar neutrinos [15] was actually preceded by evidence for oscillations
of atmospheric muon-neutrinos - most likely to tau-neutrinos [13].
Origin of atmospheric neutrinos are interactions of
cosmic rays with particles in the earth's
upper atmosphere that produce pions and muons, which subsequently
decay and emit electron- and muon-neutrinos (or antineutrinos).

Oscillation experiments are, however, only sensitive to
differences of squared masses. Hence,
the actual value of the neutrino mass
is still an open issue, and presently only the upper limit of the mass of
the antielectron-neutrino can be deduced from tritium beta decay
to be $2.2 eV/c^{2}$ [16].
>From neutrinoless double beta decay [17], a lower limit of $0.05 -
0.2eV/c^{2}$ has
been  deduced in 2002, but this result is only valid if the neutrino is its own
antiparticle, which is not certain.

Solar neutrinos are considered in section 4,
and a brief outline of some of the perspectives of the field is given
in section 5.
 \newpage
\section{Energy Evolution in Stars}

Stellar and in particular, solar energy is due to fusion of lighter nuclei
to heavier ones, which is induced by thermal motion in the star.
According to the mass formula that was derived by Weizs\"acker in 1935 [18], and
by Bethe and Bacher independently in 1936 [19], the difference in
binding energies before and after the reaction - the mass defect - is
converted to energy via Einstein's $E=mc^{2}$ [20], and is then added to
the star's energy balance.  The binding energy per nucleon rises with mass
number
starting steeply from hydrogen because the fraction of the
surface nucleons decreases, then it flattens and reaches a maximum at iron-56,
the most tightly bound nucleus;
afterwards it drops slowly towards large masses. Although this smooth
behavior of the fractional binding energy per nucleon is modified
by pairing and shell effects, the overall shape of the curve ensures that
energy can be released either by fission of heavy nuclei or by fusion of
light nuclei, as
it occurs in stars, thus providing our solar energy.

In case of main-sequence stars such as the sun there are no rapid changes in
the star that could compete with the time-scale of the nuclear reactions
and hence, the energy evolution occurs through equilibrium nuclear burning.
Most important in the solar case is hydrogen burning, where the
transformation of four hydrogen nuclei into one helium-4 nucleus
is accompanied by a mass loss of 0.71 per cent of the initial
masses, or $0.029u$. It is converted into an energy of about 26.2 MeV,
including the annihilation energy of the two positrons that are produced,
and the energy that is carried away by two electron neutrinos.
>From the known luminosity of the sun, one can calculate a total mass loss rate
of $4.25\cdot10^{9}kg/s$. At this rate, the hydrogen equivalent of one
solar mass could sustain radiation for almost $10^{11}$ years.

The reactions between nuclei inside stars are due to the thermal
motion, and are therefore called $thermonuclear$. Before stars reach an
explosive final (supernova) stage, the energy
release due to these
reactions is rather slow. From the hydrostatic equilibrium condition in
the sun one derives the central temperature as
\begin{equation}
T_{\odot}\leq\frac{8}{3}\frac{G\mu}{R}\frac{M_{\odot}}{R_\odot}.
\end{equation}
With the gas constant R, the average number of atomic mass units
per molecule $\mu$ (=0.5 for ionized hydrogen), the gravitational constant
G, the solar mass $M_{\odot}=1.99\cdot10^{30}kg$ and the solar radius
$R_{\odot}=6.96\cdot10^{8}m$ one finds the central solar temperature
\begin{equation}
T_{c}\leq3\cdot10^{7}K.
\end{equation}
Numerical solutions by Bahcall et al. [22] yield a central temperature
$T_{c}=1.57\cdot10^{7} K$ and a central pressure
$P_{c}=2.34\cdot10^{16} Pa$, with a
central solar density
of $\rho_{c}=1.53\cdot10^{5} kg/m^{3}$. For these large values of temperature,
the assumption of an ideal gas is indeed justified.
The reaction rates are strongly dependent on
temperature (typically $\sim T^{22}$ for the CNO-cycle and
$\sim T^{4}$ for the pp-chain at $T_{c}$) and therefore, massive stars have
much greater luminosities with only slightly higher central temperatures.
As was noted by Bethe already in 1938, Y Cygni has $T=3.2\cdot10^{7} K$ and a
luminosity per mass unit of $0.12 W/kg$, whereas the sun's luminosity per
mass unit
is only about $2\cdot10^{-4} W/kg$ (the most recent best-estimate value [21]
of the total solar luminosity being $3.842\cdot10^{26} W$).

Expressed in units of energy, however, the central solar temperature
is only about $1.35\cdot10^{3} keV$. This has to be compared with
the height of the Coulomb barrier
\begin{equation}
E_{coul}=\frac{Z_{1}Z_{2}e^{2}}{R}
\end{equation}
with the interaction radius R and the proton numbers $Z_{1}, Z_{2}$
of the nuclei that tend to fuse in order to release energy. Since
$E_{coul}(R)\sim Z_{1}Z_{2} MeV$, more than a factor of $10^{3}$ in thermal
energy
is missing in order to overcome the Coulomb barrier.

Thermonuclear reactions in stars can therefore only occur
due to the quantum-mechanical tunneling that was established
by Gamow [2]. The tunneling probability is
\begin{equation}
P=p_{0}E^{-1/2}exp(-2G)
\end{equation}
with the Gamow-factor
\begin{equation}
G=\sqrt{\frac{m}{2}}\frac{2 \pi Z_{1}Z_{2}e^{2}}{hE^{1/2}}.
\end{equation}
Here m is the reduced mass and $Z_{1},Z_{2}$ are the respective charges
of the fusing nuclei, and E is the energy. The factor $p_{0}$
depends only on properties of the colliding system.  For the pp-reaction at
an average
energy and at solar temperature, $P$ is of the order of $10^{-20}$.
It steeply increases with energy and decreases with the product of the
charges. Hence, at solar temperatures only systems with small
product of the charges may fuse, and for systems with larger $Z_{1}Z_{2}$
the temperature has to be larger to provide a sizeable penetration
probability. As a consequence, clearly separated stages of different
nuclear burnings occur during the evolution of a star in time.

Once the Coulomb barrier has been penetrated, an excited compound
nucleus is formed, which can afterwards decay with different probabilities into
the channels that are allowed from the conservation laws. The energy of
outgoing particles and gamma-rays is shared with the surroundings
except for neutrinos, which leave the star without interactions.

Energy levels of the decaying compound nucleus above or below the nucleon
removal
energy can be of different types, stationary levels of small width which
decay via gamma-emission, and short-lived quasi-stationary levels above the
removal energy which can also (and more rapidly) decay via particle
emission. Their width becomes larger with increasing energy and eventually
also larger than the distance between neighbouring levels.

Due to the existence of quasi-stationary levels above the nucleon removal
energy,
a compound nucleus may also be formed in a resonance when the initial energy
matches the one of an energy level in the compound nucleus. At a resonance,
the cross-section can become very large, sometimes close to the geometrical
value.
Astrophysical resonant or non-resonant cross-sections are usually written as
\begin{equation}
\sigma(E)=S\cdot E^{-1}exp(-2G)
\end{equation}
with the astrophysical cross-section factor S that contains the properties
of the corresponding reaction. Although it can be computed in principle,
laboratory measurements are a better option. However, because of the small
cross-sections, these measurements are difficult at
low energies. Extrapolations to these energies are fairly reliable
for non-resonant reactions where S(E) is a slowly varying function of E,
but this is not true in the case of resonances, which may (or may not) be hidden
in the region of extrapolation. The present state of the art for measurements
of S(E) in an underground laboratory to shield cosmic rays is shown in
figure 4 for the reaction
\begin{equation}
^{3}He(^{3}He,2p)^{4}He
\end{equation}
that is very important in the stellar pp-chain, cf. next section.
The solid line is a fit with a screening potential that accounts
for a partial shielding of the Coulomb potential of the nuclei due
to neighbouring electrons.
Data from the LUNA collaboration [23]
extend down to 21 keV, where the Gamow peak at the
solar central temperature is shown in arbitrary units. The
peak arises from the product of the Maxwell distribution at
a given temperature T and
the penetration probability. Its maximum is at an energy
\begin{equation}
E_{G}=\Bigl[\sqrt{\frac{m}{2}}\pi\frac{2 \pi Z_{1}Z_{2}e^{2}kT}{h}\Bigr]^{2/3}.
\end{equation}

At  $E_{G}$,
the S-factor for the He-3 + He-3 reaction becomes $5.3 MeVb$.
The average reaction probability per pair and second is given by
\begin{equation}
<\sigma v>=\int_{0}^{\infty}\sigma(E)vf(E)dE
\end{equation}
where f(E) can be expressed in a series expansion
near the maximum. Keeping only the quadratic terms, the reaction
probability becomes [24]
\begin{equation}
<\sigma
v>=\frac{4}{3}(\frac{2}{m})^{1/2}\frac{1}{(kT)^{1/2}}S_{G}\cdot\tau^{1/2}exp
(-\tau)
\end{equation}
with the S-factor $S_{G}$ at the Gamow peak and
\begin{equation}
\tau=3E_{G}/(kT).
\end{equation}
The temperature dependence of $<\sigma v>$ may be expressed as
\begin{equation}
\frac{\partial ln<\sigma v>}{\partial ln T}=\frac{\tau}{3}-\frac{2}{3},
\end{equation}
which can attain values  near or above 20. As a consequence of
such large values for the exponent of T, the thermonuclear
reaction rates become extremely strongly dependent on
temperature, and small fluctuations in T may
cause dramatic changes in the energy (and neutrino) production of a star.
The corresponding
uncertainty in stellar models created the long-standing
controversy about the origin of the solar neutrino deficit, which
has only recently been decided in favor of the particle-physics
explanation, cf. section 4.
\newpage
\section{Hydrogen burning}
Due to the properties of the thermonuclear reaction rates, different
fusion reactions in a star are separated by sizeable temperature
differences and during a certain phase of stellar evolution, only
few reactions occur with appreciable rates. Stellar models account
in network-calculations for all simultaneously occuring reactions.
Often the rate of the fusion process is determined by the slowest in
a chain of subsequent reactions, such as in case of the nitrogen-14
reaction of the CNO-cycle.

In hydrogen burning, four hydrogen nuclei
are fused into one helium-4 nucleus, and the mass defect of 0.71 per cent
is converted into energy (including the annihilation energy of the
two positrons, and the energy carried away by the neutrinos):
\begin{equation}
4 \cdot^{1}H\rightarrow ^{4}He + 2e^{+} + 2\nu_{e} + 26.2 MeV.
\end{equation}
As net result,
two protons are converted into
neutrons through positron emission $(beta^{+}$-decay), and because of
lepton number conservation, two electron neutrinos are emitted.
Depending on the reaction which produces the neutrinos, they can carry
between 2 and 30 per cent of the energy. Helium synthesis in stars
proceeds through different reaction chains which occur simultaneously.
The main series of reactions are the proton-proton chain, figure 2, and
the CNO-cycle, figure 3.

In the present epoch, the pp-chain
turns out to be most important
for the sun - the CNO-cycle produces
only 1.5 per cent of the luminosity [22]. The pp-chain starts with two
protons that
form a deuterium nucleus, releasing a positron and an electron neutrino.
(With much smaller probability it may also start with the p-e-p process,
figure 2). This reaction has a very small cross section, because the beta-decay
is governed by the weak interaction. At central solar temperature
and density, the mean reaction time is $10^{10}$ years, and to a certain
extent it is due to this
huge time constant that the sun is still shining. With
another proton, deuterium then reacts to form helium-3. This
process is comparably fast and hence, the abundance of deuterons in stars
is low.

To complete the chain to helium-4 three branches
are possible. The first - in the sun with
85 per cent most frequent - chain (ppI) requires two helium-3 nuclei
and hence, the first reaction has to occur twice, with two positrons and
two electron neutrinos being emitted. The other two branches
(ppII, ppIII) need helium-4 to be produced already (in previous burnings, or
primordially). In the subsequent reactions between helium-3 and helium-4,
the additional branching occurs because the product beryllium-7 can
react either with an electron to form lithium-7 plus neutrino (ppII), or
with hydrogen to form boron-8 (ppIII). The energy released by the three chains
differs because the neutrinos carry different amounts of energy with them,
and the
relative frequency of the different branches depend on temperature,
density, and chemical composition. The per centages in figure 2 refer
to the standard solar model at the present epoch [22]. Details of the
various parts of the chain including the corresponding
energy release, the energies carried away by the
neutrinos and the reaction rate constants have been discussed by Parker et
al. [26]
and Fowler et al. [27].

The other main reaction chain in hydrogen burning is the CNO-cycle,
figure 3. Here, the carbon, nitrogen and oxygen isotopes serve
as catalysts, their presence is required for the cycle to proceed.
The main cycle is completed once the initial carbon-12 is
reproduced by nitrogen-15 + hydrogen. There is also a secondary cycle (not
shown in
figure 3 since it is $10^{4}$ times less probable). It causes oxygen-16 nuclei
which are present in the stellar matter to take part in the CNO-cycle
through a transformation into nitrogen-14. The CNO-cycle
produces probably most of the nitrogen-14 found in nature. For sufficiently high
temperatures, the nuclei attain their equilibrium abundances and
hence, the slowest reaction - which is nitrogen-14 + hydrogen - determines
the time to complete the
whole circle (bottom of figure 3).

The CNO-cycle contributes only a few per cent to the luminosity of a star
with one solar mass, but it dominates in stars with masses above 1.5 times
the
solar value because its reaction rates rise much faster with
temperature as compared to pp.
Details of the Bethe-Weizs\"acker cycle
have been discussed by Caughlan and Fowler [28].
The cycle had first been proposed by Weizs\"acker in [7]. In this work,
he abandoned the main reaction path that he had considered in [6], namely,
from hydrogen via deuterium and lithium to helium, because the
intermediate nuclei of mass number 5 that were supposed
to be part of the scheme had turned out to be unstable.

In the first paper of the series [6], he had considered various reaction chains
that allow for a continuous generation of energy from the
mass defect, and also of neutrons for the buildup of
heavy elements. He had confirmed that the temperatures in the interior
of stars are sufficient to induce nuclear reactions starting
from hydrogen. In the second paper he modified the results;
in particular, he discussed the possibility that some of the
elements might have been produced before star formation by
another process.

The link between energy evolution in stars
and the formation of heavy elements as considered in [6] turned out to end up in
difficulties when calculated quantitatively. Hence, he modified his
version of the so-called "Aufbauhypothese", according to which the neutrons
necessary for the production of heavy elements should be generated
together with the energy, and decoupled the generation of energy
from the production of heavy elements. He then concluded that stellar
energy production should essentially be due to reactions between light
nuclei, with the
corresponding abundances being in agreement with observations. The
CNO-cycle was considered to be the most probable path.

In his independent
and parallel development of the CNO-cycle that was published somewhat later [9]
and contained detailed calculations, Bethe showed that "... there will
be no appreciable change in the abundance of elements heavier than
helium during the evolution of the star but only a transmutation of
hydrogen into helium. This result...is in contrast to the commonly
accepted 'Aufbauhypothese'". Here, he referred to Weizs\"acker's
first hypothesis [6] which had, however, already been modified [7].

Together with Critchfield [8], Bethe also investigated essential parts of
the pp-chain (which Weizs\"acker also mentioned) -
in particular, deuteron formation by proton combination
as the first step - and came to the conclusion that it "...gives an energy
evolution of
the right order of magnitude for the sun". Details of the pp-chain were developed
much later in the 1950s by Salpeter [29] and others. In 1938/9, however,
Bethe was
convinced that "... the reaction between two protons, while possible, is
rather slow and will therefore be much less important in ordinary
stars than the cycle (1)" namely, the CNO-cycle.

In a calculation of the
energy production by pp-chain versus CNO-cycle
(figure 5), Bethe obtained qualitatively the preponderance of H+H at low
and N+H at
high temperatures. However, the result had to be modified
in the course of time as it became evident that the pp-chain is more
important than
the CNO-cycle at solar conditions, although the Bethe-Weizs\"acker-fraction
will increase considerably in the coming 4 billion years, and
eventually supersede the contribution from the ppII-chain (figure 6).

Today, detailed solar
models allow to calculate the fractions of the solar luminosity
that are produced by different nuclear fusion reactions very precisely [22].
The model results not only agree with one another - in the neutrino flux
predictions to within about 1 per cent - they are also consistent with
precise p-mode helioseismological observations of the sun's outer
radiative zone and convective zone [30].
Moreover, the production of heavier elements up to iron in subsequent burnings
at higher temperatures [27], as well as beyond iron in the r- and s-process is
rather well-understood [31].
\section{Stellar Neutrinos}
In stellar interiors, only electron neutrinos play a role.
The interaction of neutrinos with matter is extremely small, with a
cross-section of
\begin{equation}
\sigma_{\nu}\simeq(E_{\nu}/m_{e}c^{2})^{2}\cdot 10^{-17}mb.
\end{equation}
Hence, the cross-section for neutrinos with $E_{\nu}\simeq 1 MeV$ is
$\sigma_{\nu}\simeq 3.8\cdot10^{-17}mb$, which is smaller than the
cross-section for the electromagnetic
interaction between photons and matter by a factor of about
$10^{-18}$. Associated with the cross-section is a mean free path
\begin{equation}
\lambda_{\nu}=\frac{u}{\rho\cdot\sigma_{\nu}} \simeq \frac{4 \cdot
10^{20}}{\rho}m
\end{equation}
with the atomic mass unit $u=1.66\cdot 10^{-27}kg$ and $\rho$ in $kg/m^{3}$.
In stellar matter with $\rho \simeq 1.5 \cdot 10^{3}kg/m^{3}$, the mean
free path of neutrinos is therefore approximately
\begin{equation}
\lambda_{\nu}\simeq 3\cdot 10^{17}m\simeq 10 pc \simeq 4\cdot10^{9}R_{\odot}
\end{equation}
and hence, neutrinos leave normal stars without interactions
that modify their energy. This is different
during
the collapse and supernova explosion in the final stages of the evolution
of a star where nuclear density can be reached, $\rho \simeq 2.7 \cdot
10^{17}kg/m^{3}$ such that the mean free
path for neutrinos is only several kilometers, and a transport equation for
neutrino energy has to be applied.

Here only the neutrinos from nuclear reactions in a normal main-sequence
star like the sun are considered; their
energies are (to some extent, since the continuous
distributions overlap) characteristic for specific nuclear burnings.
The pp-chain which provides most of the sun's thermonuclear energy
produces continuum neutrinos in the reactions ([32]; cf. figure 2)

\hspace{1.6cm} $^{1}H+^{1}H\rightarrow^{2}H+e^{+}+\nu_{e}\hspace{1cm}
(0.420 MeV)$

\hspace{1.6cm} $^{8}B\rightarrow^{8}Be^{*}+e^{+}+\nu_{e} \hspace{1.7cm}
(14.06 MeV)$

\hspace{1.6cm} $^{13}N\rightarrow^{13}C+e^{+}+\nu_{e}\hspace{1.7cm}
(1.20 MeV)$

\hspace{1.6cm} $^{15}O\rightarrow^{15}N+e^{+}+\nu_{e}\hspace{1.7cm}
(1.74 MeV)$

where the numbers are the maximum neutrino energies for the corresponding
reaction. In addition to these continuum neutrinos, there are neutrinos at
discrete energies from the pp-chain

\hspace{1.6cm} $^{1}H+^{1}H+e^{-}\rightarrow^{2}H+\nu_{e}\hspace{1cm}
(1.44 MeV)$

\hspace{1.6cm} $^{7}Be+e^{-}\rightarrow^{7}Li^{*}+\nu_{e}
\hspace{1.7cm}(0.861 MeV - 90 per cent)$

 \hspace{7.3cm}$(0.383 MeV - 10 per cent)$

(depending on whether lithium-7 is in the ground state, or in an exited state)

\hspace{1.8cm} $^{8}B+e^{-}\rightarrow^{8}Be+\nu_{e}\hspace{1.7cm}            (15.08 MeV).$

The CNO-cycle (figure 3) which becomes important in stars with masses above
1.5 solar masses, or in later stages of the stellar evolution (figure 8) also
produces neutrinos at discrete energies

\hspace{1.8cm} $^{13}N+e^{-}\rightarrow^{13}C+\nu_{e} \hspace{1.7cm}(2.22 MeV)$

\hspace{1.8cm} $^{15}O+e^{-}\rightarrow^{15}N+\nu_{e}\hspace{1.7cm}
(2.76 MeV).$

For experiments to detect these neutrinos when they arrive on earth 8.3
minutes after their
creation the flux at the earth's surface
is of interest. Neutrinos from the central region of the sun yield a flux
of about $10^{7}/(m^{2}\cdot s)$. The precise value as function of the
neutrino energy can be calculated from solar models ([25], figure 7). Here,
solid lines
denote the pp-chain and broken lines the CNO-cycle. The low-energy
neutrinos from the initial pp-reaction yield the largest flux. However,
the first experiment by Davis et al. [11] that detected solar neutrinos on
earth with a large-scale underground
tetrachloroethylene tank in 1967/8 - and thus confirmed the
theory how the sun shines and stars evolve -
made use of the reaction

\hspace{1.8cm} $\nu_{e}+^{37}_{17}Cl \rightarrow e^{-}+^{37}_{18}Ar-0.814 MeV$

and hence, only neutrinos with energies above 0.814 MeV
could be observed through the decay of radioactive argon nuclei - which are
mostly the solar boron-8 neutrinos,
cf. figure 7. The rate of neutrino captures is measured in solar neutrino
units; 1 SNU corresponds to $10^{-36}$ captures per second and target
nucleus. Experimental runs by Davis et al. during the 1970s, 80s and 90s
yielded a signal (after subtraction of the cosmic-ray background 1.6
kilometers underground) of $(2.3\pm0.3)$SNU, whereas the predicted capture
rates from a solar model were 0 SNU for pp (because
it is below threshold), 5.9 SNU for boron-8 beta decay, 0.2 SNU for the
pep-reaction, 1.2 SNU for beryllium-7 electron capture, 0.1 SNU for
nitrogen-13 decay and 0.4 SNU for oxygen-15 decay, totally about 8 SNU.

The observation of less than 50 per cent of the expected neutrino flux
created a controversy about the origin of
the deficit, which was finally - in 2001 - resolved [15] in favor of the
particle-physics explanation that had originally been
proposed by Pontecorvo in 1968 [33]: on their way from the solar interior
to the earth, electron-neutrinos
oscillate to different flavors which escape detection, thus
creating the deficit. Although deficiencies in the solar models could have
been responsible for the discrepancy
(in view of the sensitive dependence of the neutrino flux on the central
temperature), it could be confirmed [15] that
the models are essentially correct, giving the right value of $T_{c}$
within 1 per cent.

Before this big step in the understanding of stellar evolution and neutrino
properties could be taken, there was substantial progress both in
experimental and theoretical neutrino physics. In 1992 the
Gallex-collaboration succeeded
to measure the pp-neutrinos from the initial fusion reaction, which
contributes more than 90 per cent of the integral solar neutrino flux [12].
They used a radiochemical detector with gallium as target, exploiting the
reaction

\hspace{1.8cm} $\nu_{e}+^{71}_{31}Ga \rightarrow e^{-}+^{71}_{32}Ge - 0.23
MeV.$

The threshold is below the maximum neutrino energy for pp-neutrinos of 0.42
MeV and a large fraction of the pp-neutrinos can therefore be detected in
addition to the pep-,
beryllium-7 and boron-8 neutrinos. Gallex - which is sensitive to electron
neutrinos only - thus provided
the proof that pp-fusion is indeed the main source of solar energy. The
result $[(69.7+7.8/-8.1) SNU]$  was
confirmed by the Sage experiment $[(69\pm12) SNU]$ in the Caukasus [35].
Again this was substantially below the
range that various standard solar models predicted (120-140 SNU), and the
solar neutrino deficit persisted. At that time, there were clear
indications - but no definite evidence yet - that the flux decreases
between sun and earth due to neutrino flavor oscillations - most probably enhanced through the MSW-effect [36] in the sun -, "...pointing
towards a muon-neutrino mass of about 0.003 eV" [34]. The result was later
updated to $(73.9\pm6.2) SNU$ and could be assigned to the fundamental
low-energy neutrinos from the pp and pep reactions - but then there
remained no room to accomodate the beryllium-7 and boron-8 neutrinos.

Solar neutrinos were also detected in real time with the
Kamiokande detector [37] in Japan, a water-Cerenkov detector,
and precursor to the famous SuperKamiokande detector.
Due to the high threshold of about 7.5 MeV, it could see only the most
energetic neutrinos from the decay of boron-8 in the solar center. With the
Cerenkov light pattern one could measure for the first time the incident
direction of the scattering neutrinos, and prove that they do indeed come
from the sun. The result of the boron-8 neutrino flux was

\hspace*{1.6cm}$flux\frac{Observed}{Predicted}(\nu_{e})=0.54\pm0.07$,

again confirming the deficit. To solve the solar neutrino problems, a
larger target volume and a lower energy threshold was needed: the
SuperKamiokande detector in the same zinc mine with a threshold of 5 MeV.
Here, 32000 tons of pure water are surrounded by 11200 photomultiplier
tubes for observing electrons scattered by neutrinos (many of the tubes
were destroyed end of 2001 when one collapsed, emitting an underwater shock
wave). The detector was designed to record about 10000 solar neutrino
collisions per year - 80 times the rate of its predecessors -, but also
atmospheric neutrinos, and possible signs for proton decay. The result [14]
for the boron-8 flux  can be expressed as

\hspace*{1.6cm}$flux\frac{Observed}{Predicted}(\nu_{e})=0.47\pm0.02$,

which was in agreement with the previous findings, but more precise. There
was a massive hint that the deficit could be
due to neutrino oscillations, since the SuperKamiokande collaboration found
evidence in 1998 [13] that muon neutrinos which are produced in the upper
atmosphere by pion and muon decays change their type when they travel
distances
of the order of the earth's radius due to oscillations into another
species, most likely into tau neutrinos. The appearance of the tau
neutrinos could not yet be detected directly, but oscillations to electron
neutrinos in the given parameter range were excluded since the
$\nu_{e}$-flux was unchanged, and also by reactor data. Accelerator
experiments with a long baseline of 700 km
between neutrino source and detector are being planned to
verify this interpretation [40].

The atmospheric data showed a significant suppression of the observed
number of muon neutrinos as compared to the theoretical expectation at
large values of $x/E_{\nu}$,
with the travel distance x (large when the neutrinos travel through the
earth) and the neutrino energy $E_{\nu}$ which is in the GeV-range for
atmospheric neutrinos and hence, much higher than in the solar case.
The observed dependence on distance is expected from the
theoretical expression for oscillations into another flavor, which yields
in the model case of two flavors
\begin{equation}
P=\frac{1}{2}sin^{2}(2\theta) \cdot(1-cos(2\pi x/L)).
\end{equation}
Here, $\theta$ is the mixing angle between the two flavors considered, and
the characteristic oscillation length (the distance at which the initial
flavor content appears again) is
\begin{equation}
L=4\pi E_{\nu}/\Delta m^{2}\simeq 2.48(E_{\nu}/MeV)/(c^{4}\Delta
m^{2}/eV^{2}) [m]\end{equation}
with the difference $\Delta m^{2}=\mid m_{2}^{2}-m_{1}^{2}\mid$
of the neutrino mass eigenstates. Atmospheric neutrino experiments are thus
sensitive to differences in the squared masses of $10^{-4}$ to $10^{-2}
eV^{2}/c^{4}$ whereas solar neutrinos are sensitive to differences below
$2\cdot 10^{-4} eV^{2}/c^{4}$ due to the lower neutrino energy and the
larger distance between source and detector. Whereas such
mixings between neutral particles that carry mass had been firmly
established many years ago in the case of quarks that build up the $K^{0}$
and $B^{0}$-mesons and their antiparticles - including the proof that CP is violated [38] for three quark families -, it remained an open
question until 1998 whether the corresponding phenomenon [39] exists for
leptons.

The atmospheric SuperKamiokande data proved beyond reasonable doubt the
existence of oscillations and finite neutrino masses [13, 40], with $\Delta
m^{2}_{atm}\simeq 2.5\cdot 10^{-3}eV^{2}$ and maximal mixing
$sin^{2}(2\theta_{atm})\simeq 1$. The corresponding step for solar
neutrinos followed in 2001: charged-current results $(\nu_{e}+d\rightarrow
e+p+p$; sensitive to electron neutrinos only) from the Sudbury Neutrino
Observatory (SNO) in Canada with a $D_{2}O$-Cerenkovdetector [15], combined
with elastic scattering data from SuperK $(\nu_{\odot}+e\rightarrow \nu +
e$; sensitive to all flavors), established
oscillations of solar boron-8 neutrinos.

These results were
confirmed and improved $(5.3\sigma)$ in 2002 by neutral current results
from SNO $(\nu_{\odot}+d\rightarrow \nu+n+p)$; a further improved
measurement with salt (NaCl; chlorine-35 has a high n-capture efficiency)
is currently underway.
The total flux measured with the NC reaction is $(5.09+0.64/-0.61)\cdot
10^{6}$ neutrinos per $(cm^{2}\cdot s)$. This is in excellent agreement
with the value from solar models $(5.05+1.01/-0.81)$, proving that stellar
structure and evolution is now well-understood. The currently most-favored
mechanism for solar neutrino conversion to myon- and tauon-flavors is the
"Large mixing angle" solution, which also implies matter-enhanced
(resonant) mixing in the interior of the sun through the MSW-effect [36].

\section{Perspectives}
Since the early works by Weizs\"acker and Bethe, the investigation of
thermonuclear processes in stars
has developed into considerable detail, and with the advent of stellar
neutrino physics an independent confirmation of the origins of solar energy
has emerged.
Improvements in the precise measurements of all the reaction rates at low
energies that are involved in the fusion chains may still be expected, as
has been outlined in the model case of the $^{3}He+^{3}He$ system within
the energy region of the solar Gamow peak [23].

However, not only a good knowledge of the processes involved in equilibrium
burnings at energies far below the Coulomb barrier, but also of explosive
burning (with short-lived nuclides at energies near the Coulomb barrier) is
of interest, because both contribute to the observed abundances of the
elements. This requires new experimental facilities.

An improved understanding of the cross-sections will then put the
predictions of the solar neutrino flux and spectrum on a better basis. This
entire spectrum will be investigated with high precision in the coming
decade.
In particular, the new detector Borexino will measure the monoenergetic
beryllium-7 neutrinos at 862 keV, which depend very sensitively on the
oscillation parameters. The LENS experiment will utilize inverse beta-decay
to an isomeric state of the daughter nuclide in order to investigate the
low-energy solar neutrinos in real time with considerably reduced
background. Together with forthcoming SNO and  KamLAND results it will then
be possible to definitely determine all the mixing parameters in a
three-family scheme - and verify, or falsify, the LMA solution.

The more detailed knowledge about the physics of stars will thus be
supplemented by considerable progress regarding neutrino properties [40].
Questions to be settled are the individual neutrino masses (rather than the
difference of their squares); whether neutrinos are their own antiparticles
(to be decided from the existence or non-existence of neutrinoless double
beta-decay); whether neutrinos violate CP just as quarks do, or maybe in a
different manner that opens up a better understanding of the
matter-antimatter asymmetry of the universe than has been possible from the
investigation of quark systems (to be decided in experiments with strong
neutrino beams).

In any case, the Standard model of particle physics has to accomodate
finite neutrino masses, and in future theoretical formulations the relation between quark mixing and neutrino mixing will probably become more
transparent.
\newpage
\Large\bf
Acknowledgements\\[1.4cm]
\normalsize\rm
I am grateful to G. Schatz and W.M. Tscharnuter for corrections and
suggestions.
\newpage
\Large\bf
References\\[1.4cm]
\normalsize\rm
\hspace*{.6cm}[1] A.S. Eddington, The Internal Constitution of the Stars

\hspace*{.6cm} (Cambridge University Press, 1926).

[2] G. Gamow, Z. Physik $\bf52$, 510 (1928).

[3] R. d'E. Atkinson and F.G. Houtermans, Z. Physik $\bf54$, 656 (1929).

[4] G. Gamow and E. Teller, Phys. Rev. $\bf53$, 608 (1938).

[5] S. Chandrasekhar, An Introduction to the Study of Stellar Structure

\hspace*{.6cm}(University of Chicago Press, 1939).

[6] C.F. v. Weizs\"acker, Physik. Zeitschr.  $\bf38$, 176 (1937).

[7] C.F. v. Weizs\"acker, Physik. Zeitschr.  $\bf39$, 633 (1938).

[8] C.L. Critchfield and H.A. Bethe, Phys. Rev. $\bf54$, 248, 862 (L) (1938).

[9] H.A. Bethe, Phys. Rev. $\bf55$, 434 (1938).

[10] E.M. Burbidge, G.R. Burbidge, W.A. Fowler, F. Hoyle,

\hspace*{.6cm}Rev. Mod. Phys.$\bf29$, 547 (1957).

[11] R. Davis, D.S. Harmer, K.C. Hoffman, Phys. Rev. Lett. $\bf20$, 1205 (1968).

[12] P. Anselmann et al., Phys. Lett. B $\bf285$, 376 (1992);

\hspace*{.6cm}$\bf327$, 377 (1994); $\bf342$, 440 (1995).

[13] S. Fukuda et al., Phys. Rev. Lett. $\bf81$, 1562 (1998); $\bf85$, 3999
(2000).

[14] S. Fukuda et al., Phys. Rev. Lett. $\bf86$, 5651 (2001).

[15] Q.R. Ahmad et al., Phys. Rev. Lett. $\bf87$, 071301 (2001)

\hspace*{.6cm}and 89, 011301 (2002).

[16] C. Weinheimer et al., Phys. Lett. B $\bf460$, 219 (1999).

[17] L. Baudis al., Phys. Rev. Lett. $\bf83$, 41 (1999) and

\hspace*{.6cm}Eur. Phys. J. A$\bf12$, 147 (2001).

[18] C.F. v. Weizs\"acker, Z. Physik $\bf96$, 431 (1935).

[19] H.A. Bethe and R.F. Bacher, Rev. Mod. Phys. $\bf8$, 82 (1936).

[20] A. Einstein, Ann. Physik $\bf18$, 639 (1905).

[21] C. Fr\"ohlich and J. Lean, Geophys. Res. Lett. $\bf25$, No. 23, 4377
(1998).

[22] J.N. Bahcall, M.H. Pinsonneault and S. Basu,

\hspace*{.6cm}Astrophys. J. $\bf555$, 990 (2001).

[23] E.C. Adelberger et al., Rev. Mod. Phys. $\bf70$, 1265 (1998);

\hspace*{.6cm}M. Junker et al, LUNA collab., Nucl. Phys. Proc. Suppl.
$\bf70$, 382 (1999).

[24] R. Kippenhahn and A. Weigert, Stellar Structure and Evolution,
Springer 1990.

[25] J. Bahcall and M.H. Pinsonneault, Rev. Mod. Phys. $\bf67$, 1 (1995)

\hspace*{.6cm} and astro-ph 0010346.

[26] P.D. Parker, J.N. Bahcall and W.A. Fowler, Ap. J.  $\bf139$, 602 (1964).

[27] W.A. Fowler, G.R. Caughlan and B.A. Zimmerman,

\hspace*{.6cm} Ann. Rev. Astron. Astrophys. $\bf5$, 525 (1967).

[28] G.R. Caughlan and W.A. Fowler, Ap. J.  $\bf136$, 453 (1962).

[29] E.E. Salpeter, Phys. Rev.  $\bf88$, 547 (1952).

[30] S.A. Bludman and D.C. Kennedy, Astrophys. J.  $\bf472$, 412 (1996).

[31] K.R. Lang, Astrophysical Formulae, chapter 4 (Springer 1980),

\hspace*{.6cm}and references therein.

[32] J. N. Bahcall, Phys. Rev. $\bf135$, B 137 (1964).

[33] B. Pontecorvo, Sov. Phys. JETP $\bf26$, 984 (1968);

\hspace*{.6cm}V. Gribov and B. Pontecorvo, Phys. Lett. $\bf28$, 493 (1969).

[34] T. Kirsten in: Proc. 4th Int. Solar Neutrino Conf.,
Heidelberg,

\hspace*{.6cm}Ed. W. Hampel. MPI-HD (1997).

[35] J. Abdurashitov et al., Phys. Lett. B$\bf328$, 234 (1994).

[36] S. Mikheyev, A. Smirnov, Sov. J. Nucl. Phys. $\bf42$, 913 (1985). L.
Wolfenstein,

\hspace*{.6cm}Phys. Rev. D $\bf17$, 2369 (1978).

[37] Y. Suzuki et al., Nucl. Phys. B $\bf38$, 54 (1995).

[38] J.H. Christensen, J.W. Cronin, V.L. Fitch and R. Turlay,

\hspace*{.6cm}Phys. Rev. Lett. $\bf13$, 138 (1964).

\hspace*{.6cm}M. Kobayashi and T. Maskawa, Prog. Theor. Phys.$\bf49$, 652
(1973).

\hspace*{.6cm}H Burkhardt et al., Phys. Lett. B $\bf206$, 169 (1988).

[39] Z. Maki, N. Nakagawa and S. Sakata, Prog. Theor. Phys. $\bf28$, 870 (1962).

[40] Proc. XXth Int. Conf. on Neutrino Physics and Astrophysics, M\"unchen,

\hspace*{.6cm}Ed. F. v. Feilitzsch et al., in press (2002).
\newpage
\Large\bf
Figure captions
\normalsize\rm
\begin{description}
\item[Fig. 1.]
Solar system abundances of the nuclides relative to silicon
$(=10^{6})$ plotted as function of mass
number. The stellar nuclear processes which produce the characteristic
features are outlined. The p-p-chain and the CNO-cycle of hydrogen
burning  are discussed
in the text. Elements up to $A\leq 60$ are produced in subsequent
burnings at higher temperatures, beyond $A=60$ in supernovae
through the r-, s- and p-processes.

Source of the data:

A.G.W Cameron, Space Sci. Rev. $\bf15$, 121 (1973).

Source of the Figure:

K.R. Lang, Astrophysical Formulae, p.419. Springer (1980).
\item[Fig. 2.]
Proton-proton reactions are the main source of stellar energy in
stars with masses close to or below the solar value. They were already
briefly considered by Weizs\"acker [6] and discussed in more detail
by Critchfield and Bethe [8]. Today it is known that the
ppI branch is supplemented by the ppII branch, and the
small, very temperature-dependent ppIII branch. In the latter two branches,
additional electron neutrinos of fairly high energy are produced.
The approximate partitions refer to the sun.

Source of the Figure:

H. Karttunen et al., Fundamental Astronomy. Springer (1987).
\item[Fig. 3.]
At temperatures above 20 Million Kelvin corresponding to stars
of more than 1.5 solar masses the Bethe-Weizs\"acker-cycle is more
important than the
proton-proton chain because its reaction  rate rises faster
with temperature. This CNO-cycle was first proposed by
Weizs\"acker [7] and Bethe [9]. Here, carbon, oxygen and nitrogen act
as catalysts.

Source of the Figure:

H. Karttunen et al., Fundamental Astronomy. Springer (1987).
\newpage
\item[Fig. 4.]
The astrophysical cross-section factor S(E) for the reaction
$^{3}He(^{3}He,2p)^{4}He$.
The solid line is a fit with a screening potential.
Data from the LUNA collaboration [23]
extend down to 21 keV, where the Gamow peak at the
solar central temperature is shown in arbitrary units.

Source of the Figure:

E.C. Adelberger et al., Rev. Mod. Phys. $\bf70$, 1265 (1998).

\item[Fig. 5.]
Stellar energy production in $10^{-4}J/(kg\cdot s)$ due to the proton-
proton chain (curve H+H) and the CNO-cycle (N+H), and total
energy production (solid curve) caused by both chains.
According to this calculation by Bethe in 1938 [9], the CNO-cycle
dominates at higher than solar temperatures. Its role at and below solar
temperatures as compared to pp is, however, overestimated, cf. fig. 6.

Source of the Figure:

H.A. Bethe, Phys. Rev. $\bf55$, 434 (1938).

\item[Fig. 6.]
Fractions of the solar luminosity produced by different
nuclear fusion reactions versus solar age, with the present
age marked by an arrow (Bahcall et al. 2001 [22]). The proton-proton chain
is seen
to generate the largest luminosity fractions - in particular, through the
branch that is terminated by the $^{3}He-^{3}He$
reaction. The solid curve shows the luminosity generated
by the CNO-cycle, which increases with time, but is only a small
contribution today.

Source of the Figure:

J.N. Bahcall, M.H. Pinsonneault and S. Basu, Astrophys. J. $\bf555$, 990 (2001).
\newpage
\item[Fig. 7.]
Spectrum of solar electron neutrinos according to the
"Standard Solar Model". The largest contribution is generated
by low-energy neutrinos from the p-p chain that have been
detected by the Gallex experiment [12,34]. Solid lines indicate
neutrinos from the pp-chain, dashed lines
from the CNO-cycle. Neutrinos of higher
energies had first been observed by Davis et al. [11]. The
calculation is by Bahcall and Pinsonneault [25].
(The hep-neutrinos arise from the ppIV-reaction
$^{3}He+p\rightarrow ^{4}He+\nu_{e}+e^{+}$ which is not
shown in figure 2).

Source of the Figure:

J. Bahcall and M.H. Pinsonneault, Rev. Mod. Phys. $\bf67$, 1 (1995) and
astro-ph 0010346.

\item[Fig. 8.]
The proton-proton, beryllium-7, boron-8 and nitrogen-13
neutrino fluxes as functions of solar age, with the present
age marked by an arrow (Bahcall et al. [22]). The Standard
Solar Model ratios of the fluxes are divided by their
values at $4.57\cdot10^{9}y$, the present solar age.

Source of the Figure:

J.N. Bahcall, M.H. Pinsonneault and S. Basu, Astrophys. J. $\bf555$, 990 (2001).

[this figure should be omitted in case of space problems].

((please obtain permission to reprint the figures))
\end{description}
\end{document}